\documentclass[twocolumn,english,aps,prl,superscriptaddress]{revtex4-2}
\usepackage[latin9]{inputenc}
\usepackage{color}
\usepackage{babel}
\usepackage{amsmath}
\usepackage{amssymb}
\usepackage{graphicx}
\usepackage[pdfusetitle,
 bookmarks=true,bookmarksnumbered=false,bookmarksopen=false,
 breaklinks=false,pdfborder={0 0 1},backref=false,colorlinks=true]
 {hyperref}
\hypersetup{
 urlcolor=blue, citecolor=blue, hyperfootnotes=blue, linkcolor=blue}

\makeatletter
\usepackage{mathrsfs}
\usepackage{epsfig}
\usepackage{amsfonts}
\usepackage[figuresright]{rotating}
\usepackage{dcolumn}
\usepackage{bm}
\usepackage{color}

\makeatother

\begin{document}
\textbf{Comment on ``Time Crystal in a Single-mode Nonlinear Cavity''}

In a recent Letter \citep{PRLtoComment}, Li, Wang, Tang, and Liu
claim that a single quantum Van der Pol oscillator is a dissipative
time crystal when considering the limit of vanishing nonlinearity
(two-photon damping responsible for the saturation of the number of
excitations; $\eta\rightarrow0$ in their notation). They base their
claim on the fact that, in this limit, the Liouvillian gets infinitely-many
eigenvalues with zero real part and uniformly-spaced imaginary parts.
While indeed this is a requirement for the $t\rightarrow\infty$ state
(`steady state' in the following) to present oscillations, I argue
here that it is not enough for the system to be considered a time
crystal, as we have been promoting over the last years \citep{Li22,Li23,Ming23}.
My hope is that the discussion started here will help clarifying the
difference between self-sustained oscillations in classical limits
and true time-crystalline phases of matter, concepts that, in my opinion,
are consistently confused in the literature. While I focus on \citep{PRLtoComment},
my conclusions apply as well to other models with few bosonic modes
\citep{Seibold20,Bakker22,Cabot24} and essentially to any model in
which the thermodynamic and classical limits are not independent \citep{BTC_Iemini18,BTC_Prazeres21,BTC_Passarelli22,BTC_Mattes23}.

I base my arguments on two key points. First, note that $\eta\rightarrow0$
coincides with the classical (coherent-state) limit \citep{PRLtoComment,CNB17limitcycles}.
Hence, according to the authors' logic, any few-mode dissipative nonlinear
model would be a time crystal if it possesses self-sustained oscillatory
steady states (limit cycles) in the classical limit; this would mean
that time crystals are hardly a novel phase of matter, but just a
rebranding of a well known concept in nonlinear dynamics, definitely
not the intention of their original definition \citep{Wilczek12}.
Secondly, I show below that the oscillations of a single Van der Pol
oscillator are not robust against noise (particularly dephasing noise),
while a time crystal, as a true phase of matter that generalizes space
crystals, should be robust against all types of local spatiotemporal
fluctuations. I emphasize, however, that this does not rule out the
possibility of interesting quantum phenomena arising from spontaneous
time-translational symmetry breaking in small systems, no matter whether
they can be dubbed time crystals or not.

To illustrate the previous discussion, let us consider the master
equation of a driven Van der Pol oscillator, Eq. (1) in \citep{PRLtoComment},
but adding dephasing at a rate $\gamma$ through a term $2\gamma\mathcal{D}[\hat{a}^{\dagger}\hat{a}]\hat{\rho}$
(I adopt the notation and parameter definitions of \citep{PRLtoComment}).
Such process is ubiquitous to essentially all experimental platforms
(including the ones considered in \citep{PRLtoComment}: superconducting
circuits \citep{Blais21rev} and trapped ions \citep{Leibfried03})
and cannot be neglected in general in their long-time dynamics, specially
when any other source of phase noise is removed. At a fundamental
level, dephasing comes either from noise modulating the frequency
of the system \citep{CNB-QOnotes} (e.g., stray fields on a real or
artificial atom) or from a dispersive system-environment coupling
\citep{CarmichaelBook,GardinerZollerBook}. Both cases deal with time-local
fluctuations equivalent, for a time crystal, to the local spatial
fluctuations present in space crystals. 
\begin{figure}
\includegraphics[width=1\columnwidth]{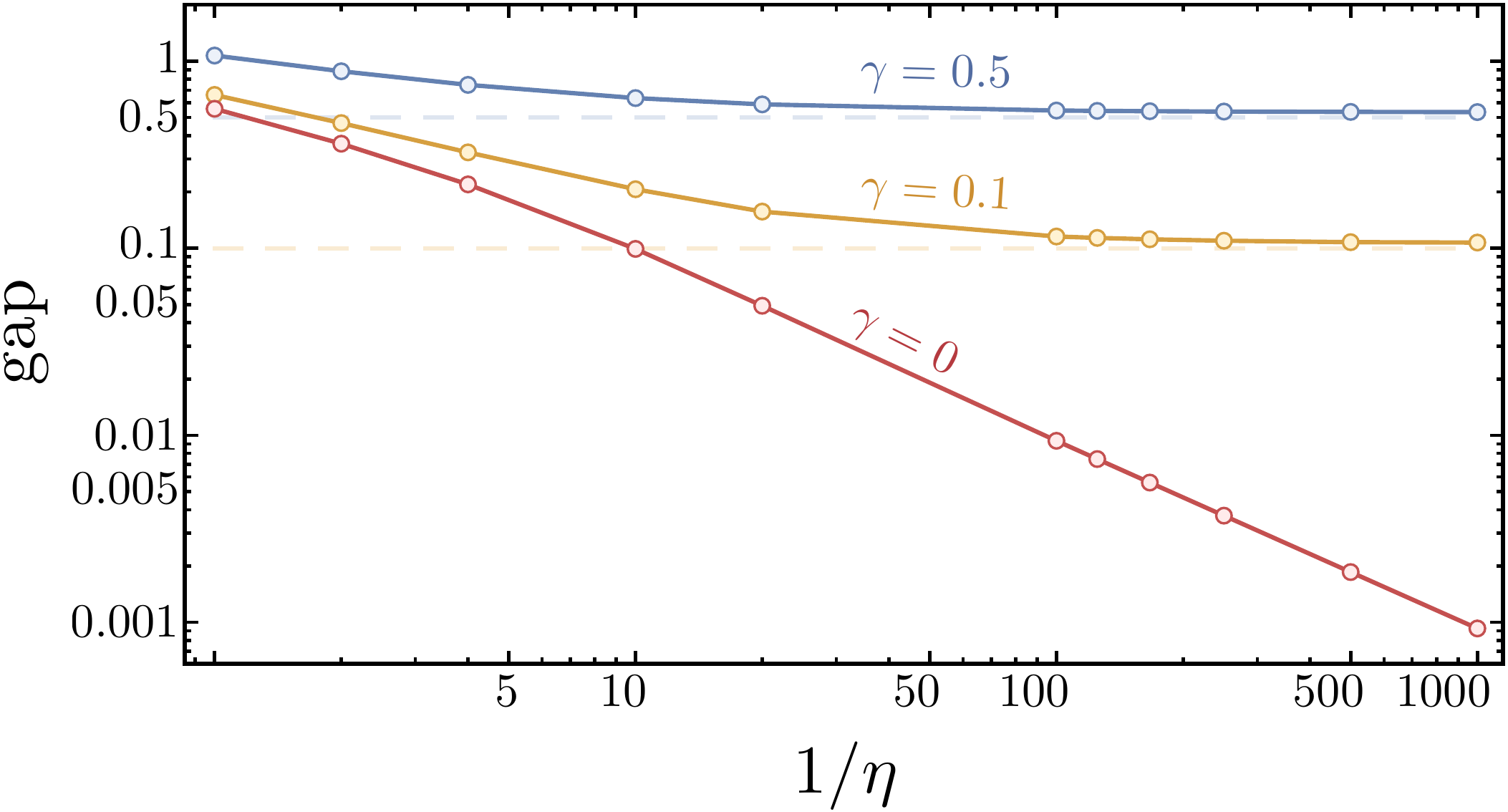} \caption{(Color online) Liouvillian gap as a function of the inverse nonlinear
loss rate $1/\eta$ for three values of the dephasing rate $\gamma$
($g=1$, $\kappa=0.1$, $\Delta=10$, and $\sqrt{\eta}\varepsilon=2$,
as in \citep{PRLtoComment}).}
\label{Fig_gap}
\end{figure}

Let us define the Liouvillian $\mathcal{L}$ as the generator of the
dynamics through $\partial_{t}\hat{\rho}=\mathcal{L}\hat{\rho}$.
For models representing real physical systems, the spectrum of this
superoperator $\mathcal{L}$ is composed of eigenvalues with non-positive
real parts. At least one eigenvalue has zero real part (defining a
subspace where steady states lie), and the gap is defined as (minus)
the smallest real part among the rest of eigenvalues. I numerically
evaluate the gap by using a sparse representation of $\mathcal{L}$
in the Fock basis \citep{CNB-NumericsNotes}, and plot it in Fig.
\ref{Fig_gap} as a function of $1/\eta$. In the absence of dephasing
($\gamma=0$), the gap goes to zero linearly with $1/\eta$, as put
forward in \citep{PRLtoComment}. However, in the presence of dephasing
($\gamma>0$), the gap saturates in the $\eta\rightarrow0$ limit
to a value close to $\gamma$. Hence, any level of dephasing will
keep the gap opened, resulting in a unique steady state with no temporal
oscillations \citep{CNB17limitcycles}.

Let us emphasize that the arguments presented here are not particular
to the Van der Pol oscillator model or the process of dephasing. The
same analysis would reach similar conclusions on other models for
which the limit of persistent oscillations coincides with the classical
limit \citep{Seibold20,Bakker22,Cabot24,BTC_Iemini18,BTC_Prazeres21,BTC_Passarelli22,BTC_Mattes23}.
Moreover, the Liouvillian gap is not only opened by dephasing, but
by other types of noises and even by any residual nonlinearity able
to make the number of excitations saturate. For example, in the case
of the Van der Pol oscillator, it is easy to check that adding a Kerr
term $\hbar U\hat{a}^{\dagger2}\hat{a}^{2}$ to the Hamiltonian leads
in the $\eta\rightarrow0$ limit to a gap on the order of $4U$.

From my point of view, the value of the models studied in \citep{PRLtoComment,CNB17limitcycles,Seibold20,Bakker22,Cabot24,BTC_Iemini18,BTC_Prazeres21,BTC_Passarelli22,BTC_Mattes23}
is that they prove how, in general, the periodic motion along limit
cycles appearing in classical dissipative systems is prone to randomly
phase drift once quantum fluctuations are considered, proving that
such cycles are not dissipative time crystals. This is reinforced
by the fact that their phase also drifts when considering other types
of fluctuations and parameter perturbations (saturating nonlinearities),
as showed in this comment.

The situation might be different for extended systems such as arrays
of Van der Pol oscillators \citep{Li22,Li23,Ming23}, which can benefit
from the many-body protection characteristic of condensed matter systems
against all kinds of local spatiotemporal fluctuations, since their
thermodynamic limit is related to the size of the array and is not
equivalent to the classical deterministic limit.

I thank Germán J. de Valcárcel, Ming Li and Eugenio Roldán for useful
discussions and suggestions, as well as funding by Generalitat Valenciana
through project CIDEXG/2023/18 and by Ministerio de Ciencia, Innovación
y Universidades through project PID2023-153363NB-C22.
\begin{flushleft}
Carlos Navarrete-Benlloch
\\
{\small\hspace*{0.2cm}Departament d'Òptica i Optometria i Ciències
de la Visió}\\
{\small\hspace*{0.2cm}Universitat de València, Spain}
\end{flushleft}

\bibliographystyle{apsrev4-1}
\bibliography{VdPnotTC_Comment}

\end{document}